\documentclass[aps,prbrapid,twocolumn,superscriptaddress,floatfix,preprintnumbers,showpacs]{revtex4-1}
\usepackage{graphicx,graphics,amssymb,amsmath,epsfig,color,subcaption,mwe,soul}
\usepackage[export]{adjustbox}

\begin{document}

\title{Non-linear and dot-dependent Zeeman splitting in GaAs/AlGaAs quantum dot arrays}
\author{V.P. Michal}
\author{T. Fujita}
\author{T.A. Baart}
\affiliation{QuTech and Kavli Institute of Nanoscience, TU Delft, 2600 GA Delft, The Netherlands}
\author{J. Danon}
\affiliation{Center for Quantum Spintronics, Department of Physics, Norwegian University of Science and Technology, NO-7491 Trondheim, Norway}
\author{C. Reichl}
\author{W. Wegscheider}
\affiliation{Solid State Physics Laboratory, ETH Z{\"u}rich, 8093 Z{\"u}rich, Switzerland}
\author{L.M.K. Vandersypen}
\affiliation{QuTech and Kavli Institute of Nanoscience, TU Delft, 2600 GA Delft, The Netherlands}
\author{Y.V. Nazarov}
\affiliation{Kavli Institute of Nanoscience, Delft University of Technology,
Lorentzweg 1, NL-2628 CJ, Delft, The Netherlands}

\begin{abstract}
We study the Zeeman splitting in lateral quantum dots that are defined in GaAs-AlGaAs heterostructures by means of split gates. 
We demonstrate a non-linear dependence of the splitting on magnetic field and its substantial variations from dot to dot and from heterostructure to heterostructure.
These phenomena are important in the context of information processing since the tunability and dot-dependence of the Zeeman splitting allow for a selective manipulation of spins.
We show that spin-orbit effects related to the GaAs band structure quantitatively explain the observed magnitude of the non-linear dependence of the Zeeman splitting. Furthermore, spin-orbit effects result in a dependence of the Zeeman splitting on predominantly the out-of-plane quantum dot confinement energy. We also show that the variations of the confinement energy due to charge disorder in the heterostructure may explain the dependence of Zeeman splitting on the dot position. This position may be varied by changing the gate voltages which leads to an electrically tunable Zeeman splitting.
\end{abstract}

\maketitle

\section{Introduction}

The realization of quantum coherent devices is an important direction in nanoscience and underlies the fields of quantum computation, information and simulation \cite{Devoret13,Hanson07,Hanson08}. The spin of an electron confined in a semiconductor quantum dot realizes an addressable and readable two-level system (qubit) with long coherence time \cite{Veldhorst15,Malinowski17}. 
Spin orbit interaction (SOI) plays a major role in semiconductor based devices and understanding its details is important for the operation of spin qubits in these systems.
In GaAs heterostructures, SOI makes possible manipulation of the spin state by time-dependent electric fields \cite{Nowack07}. It also influences the level splitting of the qubit (Zeeman splitting).
Usually the Zeeman splitting $E_Z$ is almost linear in magnetic field $B$ and can be characterized by an effective g-factor: 
\begin{equation}\label{EZ}
|g_{\rm eff}|=E_Z/\mu_B B,~B\to 0,
\end{equation}
where $\mu_B$ is the Bohr magneton.
In GaAs the bulk g-factor for electron carriers is $g_G=-0.44$. 
In addition the g-factor in nanostructures is affected by details of electron confinement \cite{Allison14,Kogan04,Potok03,Nakajima16,Noiri16}. Deviations from the bulk GaAs g-factor have been observed in various heterostructures and quantum-dot configurations, including dot-to-dot variations in quantum-dot arrays \cite{Nowack07,Koppens06,Noiri16,ShafieiThesis,Hanson03,Baart2016a,Baart17,Fujita2017}. A non-linear Zeeman splitting has also been observed, both for 2D electron gases \cite{Dobers88} and quantum dots \cite{Hanson03}.  
Apart from SOI, the observed Zeeman splitting can also be affected by dynamical nuclear polarization (see \cite{Chekhovich13} for a review).

In this article we investigate the origin of the non-linear and dot-dependent Zeeman splitting.  
We extract the Zeeman splitting from electric-dipole spin resonance (EDSR) data for a number of different quantum dots defined in AlGaAs/GaAs heterostructures (double, triple, and quadruple dot arrays) showing g-factor inhomogene{\"i}ties and g-factor variations upon changing voltages.
With a detailed analysis of the SOI in GaAs, we demonstrate that the observed non-linearity can be quantitatively explained by the interplay of SOI and vertical confinement, so that the splitting depends on the vertical confinement energy $E_{z0}$. The actual confinement energy depends on the position of the dot in the heterostructure owing to irregular placing of the donors and other sources of irregular electrostatic potential. We demonstrate that this is consistent with dot-to-dot variations of the Zeeman splitting as well as the dependence of the splitting on gate voltages.  
The understanding of the mechanism behind the dot-dependence and electrical tunability will permit efficient engineering of spin-based quantum coherent devices.

The paper is organized as follows. In Section II we summarize the theoretical analysis of SOI, we present the resulting dependence of the Zeeman splitting on confinement energy, and the non-linear Zeeman effect. In Section III we estimate the spatial variations in the splitting that can be expected from random placement of the donors. We describe experimental details in Section IV. In Section V we compare experimental and theoretical results. The details of the theoretical calculations and the full EDSR data are presented in the Supplemental Materials \cite{SM}. We finally conclude with a summary of the results and perspectives.

\begin{table*}[t]
\centering
\begin{tabular}{c c c c c c c}
  \hline\hline\noalign{\smallskip}
 Device & Ref. & $\overline{g_{\rm eff}}-g_G$ & Std. dev. of $g_{\rm eff}$ & $\kappa\,[10^{2}{\rm\,eV^{-2}}]$ & $n_0\,[10^{-3}\,{\rm nm^{-2}}]$ & Field orientation\\ \noalign{\smallskip}\hline\noalign{\smallskip}
 A double dot GaAs/Al$_{0.3}$Ga$_{0.7}$As & \cite{ShafieiThesis} & 0.0793 (0.0006) & 0.0011 & 3.0 (0.2) & 1.2 & $[110]$\\
\noalign{\smallskip}\hline\noalign{\smallskip}
B triple dot GaAs/Al$_{0.25}$Ga$_{0.75}$As & \cite{Baart2016a}, \cite{Baart17} & 0.005 (0.001) & 0.002 & 3 (1) & 2 & $[1\overline{1}0]$ \\
\noalign{\smallskip}\hline\noalign{\smallskip}
C quadruple dot GaAs/Al$_{0.307}$Ga$_{0.693}$As & \cite{Fujita2017} & 0.0054 (0.0002) & 0.0014 & 6.0 (0.1) & 2.2 & $[1\overline{1}0]$\\
\noalign{\smallskip}\hline\hline\\
\end{tabular}\\
\begin{flushleft}
\caption{The table summarizes for different devices the measured values of the effective g-factor $g_{\rm eff}$ (averaged over the ensemble of dots) relative to the bulk g-factor, its standard deviation across the measured devices, the parameter $\kappa$ that captures the non-linear Zeeman effect obtained from fitting the field-dependent data to Eq.(2), the electron concentration $n_0$ in the two-dimensional electron system and the orientation of the magnetic field.}
\label{measurements}
\end{flushleft}
\end{table*}

\section{Theoretical analysis: dependence of the Zeeman splitting on confinement energy}

Here we compute perturbatively the spin-orbit corrections to the Zeeman splitting of an electron due to confinement in a quantum dot defined in a GaAs/AlGaAs heterostructure. A magnetic field is applied in the plane of the heterostructure: ${\bf B}=(B_x,B_y,0)$. We concentrate on two special directions of the magnetic field $B_y=\pm B_x=\pm B/\sqrt{2}$, which correspond to the experimental configurations \cite{ShafieiThesis,Baart2016a,Baart17,Fujita2017}. In our coordinate system the z-axis corresponds to the heterostructure growth direction (crystal axis $[001]$) and the x-axis ($[100]$) and y-axis ($[010]$) are in the plane of the two-dimensional electron gas.

It turns out that the Zeeman splitting can be approximated with two terms that are linear and cubic in $B$: 
\begin{equation}\label{EZ}
E_Z=\mu_B B(|g_{\rm eff}|-\kappa\hbar^2\omega_L^2).
\end{equation}
Here $\omega_L=eB/2m_G$ is the Larmor frequency in the GaAs conduction band, $m_G=0.067m_0$ being the effective mass, $m_0$ is the bare mass of the electron, and $\kappa$ the parameter characterizing the non-linear Zeeman effect. Both $g_{\rm eff}$ and $\kappa$ depend on the magnetic field orientation.
The corrections to $g_{\rm eff}$ come from three leading mechanisms. The first mechanism arises from the penetration of the electron wavefunction into the AlGaAs layer with effective g-factor $g_A=0.45$ and effective mass $m_A=0.090m_0$ which both differ substantially from the values in GaAs \cite{Salis2001}. The second mechanism is due to the spin splitting of the GaAs electron spectrum in the absence of inversion symmetry, that is cubic in electron momentum\cite{Kalevich93,Falko05}. The third mechanism has been studied in \cite{Rossler,Ivchenko92} and is the momentum-dependence of the electron g-factor. With these corrections $ g_{\rm eff}$ can be expressed as
\begin{eqnarray}
&&g_{\rm eff}-g_G=g_{\rm eff}^{(1)}+g_{\rm eff}^{(2)}+g_{\rm eff}^{(3)},\label{geff1}\\
&&g_{\rm eff}^{(1)}=\frac{1}{2}(g_A-g_G)(m_A/m_G)^{1/2}(E_{z0}/\Delta)^{3/2},\label{geff2}\\
&&g_{\rm eff}^{(2)}\approx \pm 0.97(E_{z0}/E^{(2)})^{1/2},\label{geff3}\\
&&g_{\rm eff}^{(3)}\approx 1.56E_{z0}/E^{(3)}.\label{geff4}
\end{eqnarray}

The numerical factors in the expressions for $g_{\rm eff}^{(1)}$, $g_{\rm eff}^{(2)}$ and $g_{\rm eff}^{(3)}$ depend on the details of the wavefunction in the z-direction. To come to concrete values we took Airy functions corresponding to an unscreened confining electric field $\mathcal{E}$.
In the above equations $E_{z0}=(\hbar e\mathcal{E})^{2/3}/(2m_G)^{1/3}$ is the confinement energy in the z direction. It enters the corrections in a ratio with the band structure energy scales: $\Delta\approx0.3\,{\rm eV}$ is the misalignment of the edges of the conduction bands of GaAs and AlGaAs, $E^{(2)}\approx1.2\,{\rm eV}$ and $E^{(3)}\approx0.4\,{\rm eV}$ characterize the SOI. 
The $\pm$ sign in the expression for $g_{\rm eff}^{(2)}$ corresponds to $B_y=\pm B_x$.
Let us note that the corrections are proportional to different powers of $E_{z0}$ and could be expected to be of different orders of magnitude. However for the $E_{z0}$ of interest $g_{\rm eff}^{(2)}$ and $g_{\rm eff}^{(3)}$ have comparable magnitudes while $g_{\rm eff}^{(1)}$ is one order of magnitude smaller. In Eq. (\ref{geff4}) we neglected corrections of relative magnitude $\hbar\omega_0/E_{z0}\ll1$, where $\hbar\omega_0$ is the in-plane confinement energy (see Supplemental Materials \cite{SM}).

The non-linear Zeeman effect is analyzed in a similar way. We give a compact expression:
\begin{equation}\label{kappa}
\kappa\approx(\mp 0.26E_{z0}^{-3/2}+4.0E_{z0}^{-1})\,{\rm eV}^{-2}.
\end{equation}
Here $E_{z0}$ is expressed in eV.
The first term is due to the cubic spin splitting of the electron spectrum due to the absence of inversion symmetry. The second term is contributed to by the g-factor momentum dependence and Bychkov-Rashba spin splitting \cite{SM}. Again we keep here different powers of $E_{z0}$ since both terms are of comparable magnitude for the $E_{z0}$ of interest.

\section{Theoretical analysis: dot-dependent Zeeman splitting}

We have seen in the previous Section that the Zeeman splitting $E_Z$ depends on the confining energy $E_{z0}$ as a result of SOI. 
We note that $E_{z0}$ is a function of the confining field $\mathcal{E}$. As a matter of fact, the confining field is not constant over the heterostructure but fluctuates from point to point.
The reason for this is that dopants (Si atoms) in the doping layer are not uniformly distributed but their positions are random. The fluctuations of the electric field exhibit a correlation length that is approximately equal to $d$, the separation between the doping layer and the 2D gas, and have a relative amplitude
$
\delta{\mathcal E}/\mathcal{E}\sim1/2n_0^{1/2}d,
$
where $n_0$ is the 2D electron concentration.

In a typical GaAs quantum dot the electron wavefunction extends over a diameter somewhat larger than $d$. The relative variation of the field $\mathcal{E}$ over this state reads
$
\delta\mathcal{E}/\mathcal{E}=1/4(\pi n_0)^{1/2}\ell,
$
with $\ell=\sqrt{\hbar/2m_G\omega_0}$. This results in a fluctuation of the effective g-factor that is evaluated from Eqs.(\ref{geff1})-(\ref{geff4}) and reads:
\begin{equation}\label{dgeff}
\delta g_{\rm eff}=\frac{1}{4(\pi n_0)^{1/2}\ell}\Big(g_{\rm eff}^{(1)}+\frac{1}{3}g_{\rm eff}^{(2)}+\frac{2}{3}g_{\rm eff}^{(3)}\Big).
\end{equation}

Changing the gate voltages displaces the dot and changes its shape, both of which affect the expectation value of $\mathcal{E}$ and hence $g_{\rm eff}$. We explain the gate tunability of $E_{Z}$ by these effects. The maximum variation of $g_{\rm eff}$ by gate voltages occurs when a dot is displaced by a full correlation length of the electric field, in which case $g_{\rm eff}$ changes by the order of $\delta g_{\rm eff}$.

\begin{figure}[!b]
\centering
\includegraphics[width=8cm]{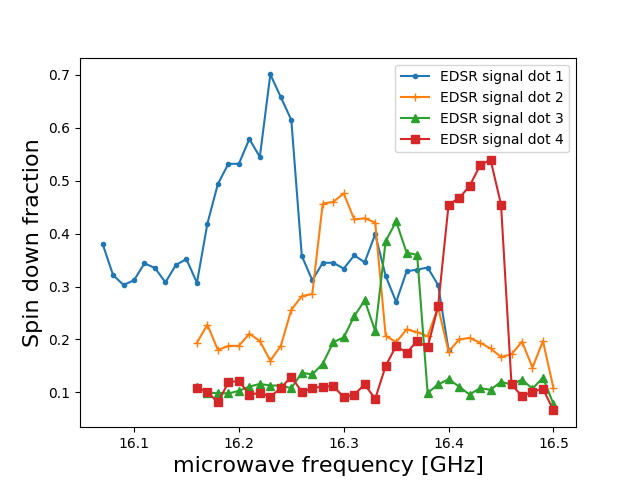}
\caption{Electric-dipole spin resonance response in a quadruple quantum dot, measured using adiabatic rapid passage to invert the spin, as in \cite{Shafiei2013}. The spin-down probability as a function of the microwave carrier frequency is shown at magnetic field $B=2.7$ T. The resonance frequency is taken to be the high-frequency edge of the resonance peak subtracted by the half-width of the frequency modulation applied for the spin inversion technique.}
\label{fig1}
\end{figure}

\begin{figure}[!b]
\centering
\includegraphics[width=8cm]{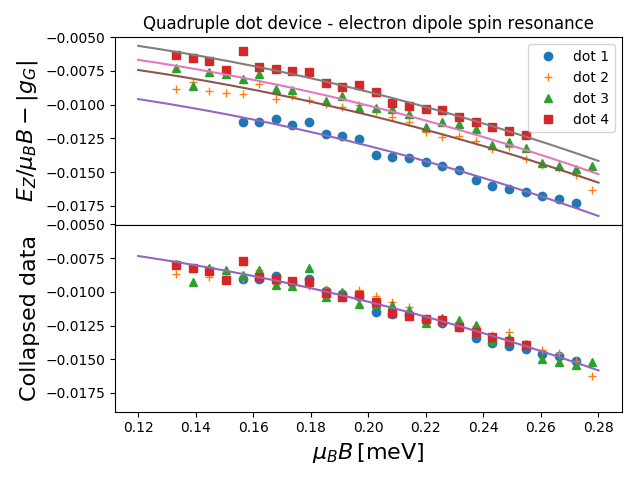}
\caption{Magnetic field dependence of the Zeeman splitting for the four dots in the quadruple dot device of Fig. \ref{fig1}. The top panel shows the bare data showing different Zeeman splittings for the four different dots. The bottom panel shows the collapsed data obtained by subtracting $(g_i-\overline{g})\mu_B B$ from the resonance energy $E_Z$ of dot $i$, where $\overline{g}$ is the mean g-factor of the four dots.}
\label{fig2}
\end{figure}

\begin{figure}[h!]
\centering
\includegraphics[width=8cm]{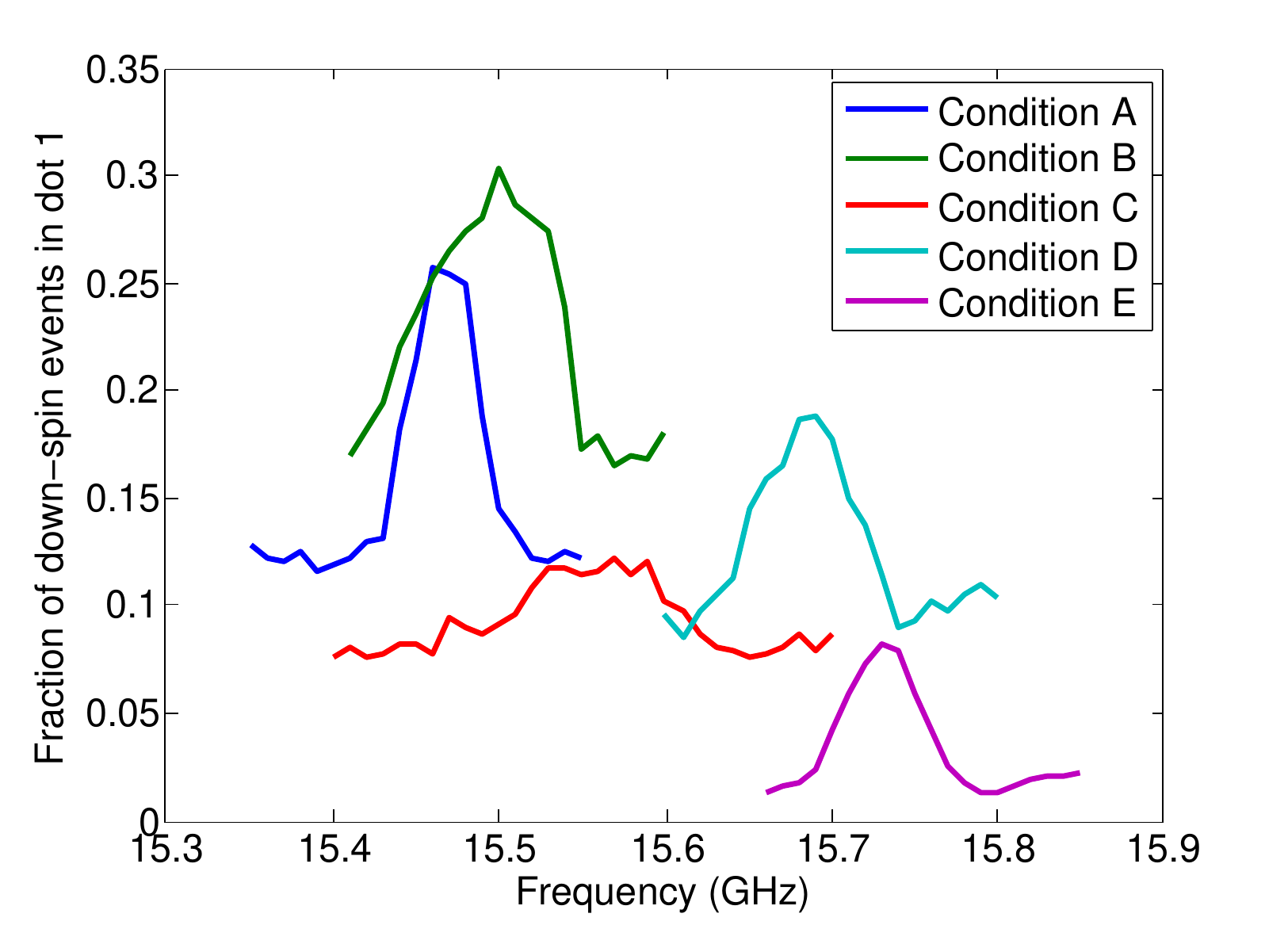}
\caption{EDSR measurement on dot~1 in the quadruple quantum dot sample (Table \ref{measurements}) in a magnetic field of magnitude $B=2.5\,{\rm T}$. Each condition corresponds to a gate voltage condition listed in Table~III in the Supplemental Materials. We observe up to 300~MHz difference in the resonance frequencies when tuning gate voltages by up to 20~mV. Resonance peaks are predominantly broadened by the 44~MHz frequency modulation applied for adiabatic spin inversion.}
\label{fig3}
\end{figure}

\section{Experiment}

The Zeeman splitting energies are measured by searching for a resonant response in EDSR experiments. 
Figure~\ref{fig1} shows as an example a measurement from a quadruple quantum dot device (see \cite{Baart2016a,Shafiei2013} for details, the measurements on this sample were reported in \cite{Fujita2017}). Data for this and two other devices are summarized in Table~\ref{measurements}.
We see that the dot-to-dot variations of the g-factor in the same sample are about 1~\% of the GaAs bulk value of $-0.44$. 

We plot in Fig.~\ref{fig2} $E_Z$ for the four different dots as a function of the magnetic field. The solid lines correspond to fits with Eq.\ref{EZ}. We observe that the data collapse (lower part of Fig.~\ref{fig2}) upon shifting $g_{\rm eff}$ for each dot individually, which means that there is no dot-to-dot dependence seen in $\kappa$. This is consistent with the fact that the possible variations of $\kappa$ are of the same order of magnitude as the experimental uncertainties.

We observe that tuning the dots by gate voltages changes the Zeeman splitting. Figure~\ref{fig3} shows resonant responses measured in dot 1 of the quadruple dot sample for a set of five different gate voltage conditions \cite{SM}.  We observe up to 300~MHz differences in the resonance frequency when changing the gate voltages up to 20~mV. A modest change in gate voltage can thus lead to a 2$\%$ change in g-factor, indicating significant electrical tunability.
In the Supplemental Materials we show additional measurements for the triple and quadruple dot devices. The absence of hysteresis and the weak time-dependence of the splitting indicate that the observed shifts of the resonance lines are unlikely due to magnetic field variations caused by the background nuclear spins.

\section{Comparison between theory and experiment}

Let us compare our predictions of Eqs.(\ref{geff1})-(\ref{dgeff}) with the measurement results of Table \ref{measurements}.
We estimate the confining energies for the structures A,B,C as $E_{z0}=12,16$, and $18\,{\rm meV}$ respectively. Depending on the magnetic field orientation, the contributions from the dominant terms $g_{\rm eff}^{(2)}$ and $g_{\rm eff}^{(3)}$, add up ($B$ along $[110]$) or subtract from each other ($B$ along $[1\overline{1}0]$). Consistent with this prediction, the experimentally measured deviation from $g_G$ is ten times bigger for device A than for device B and C.
Furthermore, in this confinement energy range both $g_{\rm eff}^{(2)}$ and $g_{\rm eff}^{(3)}$ are close to the maximum $g_{\rm eff}-g_G\approx0.08$ measured in structure A. For structure C we compute $g_{\rm eff}^{(2)}\approx\pm0.12$, and $g_{\rm eff}^{(3)}\approx0.07$, yielding a partial cancellation of their contributions. In the experiment, an even smaller residual value for $g_{\rm eff}-g_G$ is observed (a similar observation applies to device B).
We note that a very good agreement is obtained if $g_{\rm eff}^{(2)}$ is roughly half the predicted value. Then for the $[1\overline{1}0]$ orientation the terms $g_{\rm eff}^{(2)}$ and $g_{\rm eff}^{(3)}$ almost exactly compensate each other resulting in $g_{\rm eff}-g_G$ that is one order of magnitude smaller than for the other direction. In this situation, the smaller term $g_{\rm eff}^{(1)}\approx0.007$ becomes significant.
We note that the concrete values of the numerical coefficients in Eqs.(\ref{geff2})-(\ref{geff4}) do depend on the details of the confinement in the z-direction. We use a simple model for confinement which disregards the screening of the confining field by 2D carriers. Accounting for screening would generally reduce these coefficients. 

Eq.(\ref{dgeff}) gives g-factor fluctuations $\delta g_{\rm eff}\approx10^{-3}$ and $\delta g_{\rm eff}\approx2\times10^{-3}$ for the structures B and C. This agrees with the experimental values $\delta g_{\rm eff}\sim10^{-3}-2\times10^{-3}$. For the structure A, where the magnetic field orientation is different and g-factor variations can be larger, we predict $\delta g_{\rm eff}=10^{-2}$. This is larger than the observed g-factor difference for the two dots in this device, but as noted before $\delta g_{\rm eff}$ is the typical variation and it is possible that larger differences would be observed for a multi-dot device.
 
Finally, in agreement with the experimental observations we predict the parameter of the non-linear Zeeman effect $\kappa$ to be positive for the two orientations of the magnetic field. For device B,
Eq.(\ref{kappa}) is evaluated as $\kappa\approx3.5\times10^{2}\,{\rm eV}^{-2}$ in excellent agreement with the measurement. For the other structures, the prediction for $\kappa$ agrees within a factor of 2 to 3.

\section{Conclusion}

We have studied the Zeeman splitting in lateral quantum dots defined by electrostatic gates in GaAs-AlGaAs heterostructures. 
We studied the non-linear field dependence of the splitting, its spatial variations within a given structure and its changes from structure to structure.
We evaluated the spin-orbit interaction effects and found that they can explain the observed non-linear Zeeman splitting and the dependence of splitting on confinement energy.
As a consequence, the variations of confinement energy due to charge disorder in the heterostructure may explain the dependence of Zeeman splitting on the dot position. This position can be varied by changing the gate voltages and provide tunability of the Zeeman splitting.
These observations are important for quantum information processing since the Zeeman splitting differences enable site-selective manipulation.

\emph{Acknowledgements.} This work was partly supported by the Research Council of Norway through its Centres of Excellence funding scheme, project number XXXXX, “QuSpin”, the European Research Council (ERC-Synergy), the Intelligence Advanced
Research Projects Activity (IARPA) Multi-Qubit Coherent Operations (MQCO)
Program, the Japan Society for the Promotion of Science (JSPS)
Postdoctoral Fellowship for Research Abroad, and the Swiss National
Science Foundation. We also acknowledge helpful discussions with Mark Rudner and Bertrand Halperin.

\clearpage

\widetext
\begin{center}
\textbf{\large Supplemental Materials: Non-linear and dot-dependent Zeeman splitting in GaAs/AlGaAs quantum dot arrays}
\end{center}
\setcounter{equation}{0}
\setcounter{figure}{0}
\setcounter{table}{0}
\setcounter{page}{1}
\setcounter{section}{0}
\makeatletter
\renewcommand{\theequation}{S\arabic{equation}}
\renewcommand{\thefigure}{S\arabic{figure}}
\renewcommand{\bibnumfmt}[1]{[S#1]}
\renewcommand{\citenumfont}[1]{S#1}

Here we first present the model that we use to describe the spin-orbit interaction for the confined electron and we derive the corresponding Zeeman splitting corrections and inhomogeneities due to charge disorder (part I). In part II we give the values of the voltages applied on gate electrodes for different tuning conditions. In part III we show the EDSR data for the different setups (bare and collapsed data) in a similar fashion as for the quadruple dot device in Fig. 2 of the main text. In part IV we expose the observed time evolution of the resonance signals due to nuclear spins and we argue why nuclear spin effects are negligible in our case.
    
\section{Model}

The effective mass Hamiltonian for the electron with spin-orbit coupling up to second-order in $k\cdot p$ perturbation theory writes:
\begin{equation}\label{H0}
H_0=\sum_{\alpha=x,y,z}\Big(\frac{{\hat p}_\alpha^2}{2m}+\frac{g}{2}\mu_B \sigma_\alpha B_\alpha\Big)+U(x,y,z).
\end{equation}
Here $B_\alpha$ are the vector components of the applied magnetic field, $\sigma_\alpha$ the Pauli matrices, ${\hat p}_\alpha=p_{\alpha}+eA_\alpha({\bf r})$ is the gauge-invariant momentum, $e>0$ is the elementary charge, $\mu_B=\hbar e/2m_0$ is the Bohr magneton, $m_0$ is the electron bare mass, $m,g$ are the medium-dependent electron effective mass and g-factor, and $U(x,y,z)$ is the gate defined confining potential.
We choose our coordinate system such that the z-axis corresponds to the heterostructure growth direction (crystal axis $[001]$) and the x-axis ($[100]$) and y-axis ($[010]$) define the heterostructure plane.

The confining potential consists of the triangular potential along the z direction and the gate-defined parabolic in-plane potential. Neglecting the disorder potential due to randomly distributed donors, dot potential energy for $z>0$ (GaAs) reads
\begin{equation}
 U(x,y,z)=\frac{1}{2}m_G\omega_0^2(x^2+y^2)+e\mathcal{E}z.
\end{equation}
Here ${\mathcal E}=en_0/\varepsilon_G\varepsilon_0$, $\varepsilon_G\approx12.9$ being the  dielectric constant of GaAs, $\varepsilon_0$ the vacuum permittivity, and $n_0$ the electron planar concentration in the heterostructure inversion layer. This expression for the electric field close to the interface corresponds to the condition for the 2D electron gas $\mathcal{E}(z)\to0$ as $z\to\infty$. On the other hand $U(x,y,z)=\Delta\approx 300\,{\rm meV}$ for $z<0$ (AlGaAs).
The system-specific parameters for the AlGaAs-GaAs heterostructure are given in Table \ref{parameters_hs}.
\begin{table}[h!]
\begin{center}
\begin{tabular}{c c c}
  \hline\hline\noalign{\smallskip}
  Parameter & GaAs & Al$_{0.3}$Ga$_{0.7}$As\\ \noalign{\smallskip}\hline\noalign{\smallskip}
  $m$ & $m_G=0.067m_0$ & $m_A=0.090m_0$\\ \noalign{\smallskip}\hline\noalign{\smallskip}
  $g$ & $g_G=-0.44$ & $g_A=0.45$\\ \noalign{\smallskip}\hline\hline\\
\end{tabular}\\
\end{center}
\caption{Effectives masses and bulk g-factors of GaAs and Al$_{0.3}$Ga$_{0.7}$As \cite{S_Rossler}.} \label{parameters_hs}
\end{table}

At zero magnetic field the z-component of the wave function is $\psi_{z,n}(z)=c_<\exp(z\sqrt{2m_A(\Delta-E_{z,n})}/\hbar)$ if $z<0$ and $\psi_{z,n}(z)=c_>{\rm Ai}(z/z_0-E_{z,n}/E_{z0})$ if $z>0$, where ${\rm Ai}$ is the Airy function, and
\begin{equation}
E_{z0}=\frac{(\hbar e\mathcal{E})^{2/3}}{(2m_G)^{1/3}}=\frac{\hbar^2}{2m_Gz_0^2}.
\end{equation}
Boundary conditions $\psi_{z,n}(0-)=\psi_{z,n}(0+)$ and $\psi_{z,n}'(0-)/m_A=\psi_{z,n}'(0+)/m_G$ lead to energy levels $E_{z,n}=E_{z0}(|a_{n}|-\sqrt{m_AE_{z0}/m_G\Delta})$
in the regime $E_{z0}/\Delta\ll1$, where $a_{n}$ are the Airy function zeros ($n\geq 1$; for the ground state: $a_1\approx-2.34$).

To take spin-orbit into account Eq.(\ref{H0}) should be supplemented with six other terms describing the spin splitting \cite{S_Winkler03}:
\begin{equation}\label{H1}
H_1=\sum_{\iota}H_1^{(\iota)},~~H_1^{(\iota)}=\gamma_{\iota}\sum_{\alpha=x,y,z}\sigma_\alpha{\mathcal K}_\alpha^{(\iota)*},
\end{equation}
that are derived up to fourth order in $k\cdot p$ perturbation theory. For in-plane magnetic field $\mathbf{B}=(B_x,B_y,0)$ the relevant irreducible tensor components of the point group $T_d$ read \cite{S_Winkler03}:
\begin{eqnarray}\label{invariants}
&&\nonumber\mathcal{K}^{(2)*}=\frac{1}{2\hbar^3}(\{{\hat p}_x,{\hat p}_y^2-{\hat p}_z^2\},\{{\hat p}_y,{\hat p}_z^2-{\hat p}_x^2\},\{{\hat p}_z,{\hat p}_x^2-{\hat p}_y^2\})\\
&&\nonumber\mathcal{K}^{(3)*}=\frac{e}{\hbar^3}({\hat p}_x^2+{\hat p}_y^2+{\hat p}_z^2)(B_x,B_y,0)\\
&&\nonumber\mathcal{K}^{(4)*}=\frac{e}{2\hbar^3}(\{{\hat p}_x,{\hat p}_y\}B_y,\{{\hat p}_y,{\hat p}_x\}B_x,\{{\hat p}_z,{\hat p}_x\}B_x+\{{\hat p}_z,{\hat p}_y\}B_y)\\
&&\nonumber\mathcal{K}^{(5)*}=\frac{e}{\hbar^3}({\hat p}_x^2B_x,{\hat p}_y^2B_y,0)\\
&&\nonumber\mathcal{K}^{(6)*}=\frac{e\mathcal{E}}{\hbar}({\hat p}_y,-{\hat p}_x,0)\\
&&\nonumber\mathcal{K}^{(7)*}=\frac{e^2\mathcal{E}}{\hbar}(B_y,B_x,0)
\end{eqnarray}
Here $\{a,b\}=ab+ba$ and $e\mathcal{E}=\partial_zU$ is the potential energy gradient along the crystal growth direction at the heterostructure interface \cite{S_commentRashba}. The coefficients of the invariant decomposition found in literature for GaAs \cite{S_Winkler03,S_Rossler} are given in Table \ref{parameters_so}.

\begin{table}[h!]
\begin{center}
\begin{tabular}{c c c c c c}
  \hline\hline\noalign{\smallskip}
  $\gamma_{2}$ & $\gamma_{3}$ & $\gamma_{4}$ & $\gamma_{5}$ & $\gamma_{6}$ & $\gamma_{7}$\\ \noalign{\smallskip}\hline\noalign{\smallskip}
  27.6 ${\rm eV\cdot\AA^3}$ & 493 ${\rm eV\cdot\AA^4}$& -433 ${\rm eV\cdot\AA^4}$& 58 ${\rm eV\cdot\AA^4}$& 5.5 ${\rm\AA^2}$& -5.2 ${\rm\AA^3}$\\ \noalign{\smallskip}\hline\hline\\
\end{tabular}\\
\end{center}
\caption{Spin-orbit coupling parameters of GaAs \cite{S_Winkler03,S_Rossler}.} \label{parameters_so}
\end{table}

We choose the gauge where the potential vector components write $A_x=B_y(z-a),\,A_y=-B_x(z-a)$ and the canonical momenta are
$$
{\hat p}_x=p_x+eB_y{\tilde z},~~{\hat p}_y=p_y-eB_x{\tilde z},~~{\hat p}_z=p_z,~~\tilde{z}=z-a.
$$
Hence we solve the problem by the method of perturbation: the simply computable part is $H=\sum_\alpha(p_\alpha^2/2m+g\mu_B \sigma_\alpha B_\alpha/2)+U(x,y,z)$, and the perturbation includes the influence of the magnetic field on the electron motion and spin-orbit interaction: $H'=e^2B^2{\tilde z}^2/2m+e{\tilde z}(p_xB_y-p_yB_x)/m+H_1$.
Parameter $a$ is fixed by the requirement of vanishing z-averaged group velocity at zero momentum in the $x$ and $y$ directions: $v_x,\,v_y\rightarrow0$ as $p_x,\,p_y\rightarrow0$. This corresponds to the only condition at second order of perturbation theory:
\begin{equation}
\langle{\tilde z}\rangle+\frac{e^2B^2}{m}\sum_{n\neq 1}\frac{\langle1|{\tilde z}|n\rangle\langle n|{\tilde z}^2|1\rangle}{E_{z,1}-E_{z,n}}=0,
\end{equation}
where $|n\rangle$ is the eigenstate of the z-dependent part of $H$ with energy $E_{z,n}$, and $\langle\cdot\rangle$ stands for averaging over the ground state ($n=1$). Therefore up to second order in magnetic field $a$ has the explicit expression:
\begin{equation}
a=\langle z\rangle-\frac{e^2B^2}{mE_{z0}}\sum_{n\neq 1}\frac{\langle 1|z|n\rangle\langle n|(z-\langle z\rangle)^2|1\rangle}{|a_n|-|a_1|}.
\end{equation}
Here $\langle z\rangle=\frac{2}{3}|a_1|z_0\approx1.56z_0$ and we numerically compute 
$
\sum_{n\neq 1}\frac{\langle 1|z|n\rangle\langle n|(z-\langle z\rangle)^2|1\rangle}{|a_n|-|a_1|}\approx0.16z_0^3,
$
giving
$
a/z_0\approx1.56-0.32\hbar^2\omega_L^2/E_{z0}^2,
$
where $\omega_L=eB/2m_G$ is the GaAs conduction electron Larmor frequency.

\subsection{Evaluation of Zeeman splitting corrections}

The first mechanism of the g-factor deviation with respect to the GaAs bulk value comes from averaging $g$ in Eq.(\ref{H0}) for the ground state wavefunction taking into account the wave-function penetration in the AlGaAs layer. Because the g-factors of the two materials have opposite signs, one may expect a significant effect. We find that the averaged g-factor reads:

\begin{equation}\label{dg0}
\langle g\rangle=g_G+\frac{1}{2}(g_A-g_G)(m_A/m_G)^{1/2}(E_{z0}/\Delta)^{3/2}.
\end{equation}
The higher-order correction to the effective g-factor that is quadratic in $B$ is of the order $(\mu_B B/E_{z0})^2(E_{z0}/\Delta)^{7/2}$.

The second mechanism, arising from bulk inversion asymmetry (term (2) in Eq.(\ref{H1}), also known as the Dresselhaus spin-orbit term) leads to a dependence of the Zeeman splitting on the in-plane field orientation \cite{S_Kalevich93,S_Falko05}. 
Assuming $E_{z0}\gg \hbar\omega_0$ the energy correction at first-order of perturbation theory is: 
\begin{equation}
\langle H_1^{(2)}\rangle=-\frac{\gamma_2}{\hbar^3}\big(eB_y\sigma_x(\langle p_z{\tilde z}p_z\rangle-e^2B_x^2\langle \tilde{z}^3\rangle)+(x\leftrightarrow y)\big).
\end{equation}
We compute $\langle p_z{\tilde z}p_z\rangle=(\hbar^2/z_0)(-4a_1^2/45+C\hbar^2\omega_L^2/E_{z0}^2$), with $4a_1^2/45\approx0.49$, $C\approx0.25$, and $\langle(z-\langle z\rangle)^3\rangle\approx0.21z_0^3$, leading to
\begin{equation}
\langle H_1^{(2)}\rangle\approx(2m_0\gamma_2/\hbar^2z_0)(0.49-0.25\hbar^2\omega_L^2/E_{z0}^2+0.21z_0^4e^2B_x^2/\hbar^2)\mu_BB_y\sigma_x+(x\leftrightarrow y).
\end{equation}
Defining $E^{(2)}=\hbar^6/8m_0^2m_G\gamma_2^2\approx 1.2\, {\rm eV}$, this becomes in the special case $B_x=\pm B_y$:
\begin{equation}\label{H12}
\langle H_1^{(2)}\rangle\approx\Big(\frac{E_{z0}}{E^{(2)}}\Big)^{1/2}\Big(0.49-0.14\frac{\hbar^2\omega_L^2}{E_{z0}^2}\Big)\mu_B(B_y\sigma_x+B_x\sigma_y).
\end{equation}
The second order correction in $H'$ of third power in magnetic field is
\begin{equation}
\delta\langle H_1^{(2)}\rangle=-\frac{e^3B^2\gamma_2}{\hbar^3m}\sum_{n\neq 1}\frac{\langle1|{\tilde z}^2|n\rangle\langle n|p_z{\tilde z}p_z|1\rangle}{E_{z,1}-E_{z,n}}(B_y\sigma_x+B_x\sigma_y)\approx0.008\Big(\frac{E_{z0}}{E^{(2)}}\Big)^{1/2}\frac{\hbar^2\omega_L^2}{E_{z0}^2}\mu_B(B_y\sigma_x+B_x\sigma_y).
\end{equation}

The third mechanism, terms (3), (4) and (5) appear at fourth order in $k\cdot p$ perturbation theory \cite{S_Rossler}. To first order in $H'$ and in the regime $E_{z0}\gg\hbar\omega_0$, term (3) evaluates as
\begin{equation}\label{H13}
\langle H_1^{(3)}\rangle=\frac{e\gamma_3}{\hbar^3}\Big(\frac{|a_1|\hbar^2}{3z_0^2}+\frac{4}{45}a_1^2z_0^2e^2B^2\Big)(B_x\sigma_x+B_y\sigma_y)\approx\frac{E_{z0}}{E^{(3)}}\Big(0.78+0.49\frac{\hbar^2\omega_L^2}{E_{z0}^2}\Big)\mu_B(B_x\sigma_x+B_y\sigma_y),
\end{equation}
with $E^{(3)}=\hbar^4/4m_0m_G\gamma_3\approx0.43\,{\rm eV}$. Taking into account the in-plane confinement energy leads to a correction to Zeeman splitting equal to $\hbar\omega_0/E^{(3)}$, negligible when $\hbar\omega_0/E_{z0}\ll1$. The second order correction is
\begin{equation}
\delta\langle H_1^{(3)}\rangle=\frac{e^3B^2\gamma_3}{\hbar^3m}\sum_{n\neq 1}\frac{\langle1|{\tilde z}^2|n\rangle\langle n|p_z^2|1\rangle}{E_{z,1}-E_{z,n}}(B_x\sigma_x+B_y\sigma_y)\approx0.32\frac{\hbar^2\omega_L^2}{E^{(3)}E_{z0}}\mu_B(B_x\sigma_x+B_y\sigma_y).
\end{equation}
Term (4) develops as ($B_x=\pm B_y$):
\begin{equation}
\langle H_1^{(4)}\rangle=-\frac{e^3\gamma_4}{\hbar^3}\langle{\tilde z}^2\rangle B_y^2 B_x\sigma_x+(x\leftrightarrow y)\approx0.24\frac{\hbar^2\omega_L^2}{E^{(4)}E_{z0}}\mu_B(B_x\sigma_x+B_y\sigma_y),
\end{equation}
with $E^{(4)}=\hbar^4/4m_0m_G|\gamma_4|\approx0.5\,{\rm eV}$, and term (5) as:
\begin{equation}
\langle H_1^{(5)}\rangle=\frac{e^3\gamma_5}{\hbar^3}\langle{\tilde z}^2\rangle B_y^2 B_x\sigma_x+(x\leftrightarrow y)\approx0.24\frac{\hbar^2\omega_L^2}{E^{(5)}E_{z0}}\mu_B(B_x\sigma_x+B_y\sigma_y),
\end{equation}
with $E^{(5)}=\hbar^4/4m_0m_G\gamma_5\approx3.7\,{\rm eV}$.
Moreover structural inversion asymmetry induces an interaction (6) (Bychkov-Rashba spin-orbit coupling) which leads to
\begin{equation}
\langle H_1^{(6)}\rangle=-\frac{\gamma_6}{\hbar}e^2\langle\mathcal{E}{\tilde z}\rangle(B_x\sigma_x+B_y\sigma_y)\approx-0.32\frac{\hbar^2\omega_L^2}{E^{(6)}E_{z0}}\mu_B(B_x\sigma_x+B_y\sigma_y),
\end{equation}
with $E^{(6)}=\hbar^2/2m_0\gamma_6\approx0.73\,{\rm eV}$. The last invariant (7) gives
\begin{equation}
 \langle H_1^{(7)}\rangle=-\left(\frac{E_{z0}}{E^{(7)}}\right)^{3/2}\mu_B(B_y\sigma_x+B_x\sigma_y),
\end{equation}
where $E^{(7)}=\hbar^2/(2m_0|\gamma_7|)^{2/3}(2m_G)^{1/3}\approx3.13\,{\rm eV}$. This term can therefore be neglected.

As a result spin-orbit theory predicts a Zeeman energy splitting of the form
\begin{equation}
E_Z=\mu_B B(|g_{\rm eff}|-\kappa\hbar^2\omega_L^2).
\end{equation}
The effective g-factor can be written as:
\begin{equation}
g_{\rm eff}-g_G=g_{\rm eff}^{(1)}+g_{\rm eff}^{(2)}+g_{\rm eff}^{(3)},
\end{equation}
where corrections $g_{\rm eff}^{(1)}$, $g_{\rm eff}^{(2)}$, and $g_{\rm eff}^{(3)}$ correspond to wavefunction penetration in the AlGaAs material, Dresselhaus bulk inversion asymmetry spin-orbit coupling, and fourth-order in momentum dispersion respectively, with expressions:
\begin{equation}
g_{\rm eff}^{(1)}=\frac{1}{2}(g_A-g_G)(m_A/m_G)^{1/2}(E_{z0}/\Delta)^{3/2},~~
g_{\rm eff}^{(2)}=\pm\frac{8a_1^2}{45}(E_{z0}/E^{(2)})^{1/2},~~
g_{\rm eff}^{(3)}=\frac{2|a_1|}{3}E_{z0}/E^{(3)}.
\end{equation}
In $g_{\rm eff}^{(2)}$ the positive sign corresponds to $B_x=B_y$ and the negative sign to $B_x=-B_y$. 
When $E_{z0}$ is expressed in eV this becomes:
\begin{equation}\label{geff}
 g_{\rm eff}-g_{G}\approx 3.1E_{z0}^{3/2}\pm0.89E_{z0}^{1/2}+3.6E_{z0},
\end{equation}
and in addition the non-linear Zeeman effect parameter evaluates to:
\begin{equation}\label{deznl}
 \kappa\approx(\mp 0.26E_{z0}^{-3/2}+4.0E_{z0}^{-1})\,{\rm eV}^{-2}.
\end{equation}

\subsection{Influence of potential inhomogeneity} 

Electrostatic potential fluctuations due to randomly distributed donors at distance $d$ from the doping plane read
\begin{equation}
\delta \varphi({\bf r},d)=-\frac{e}{4\pi\varepsilon_0\varepsilon}\int d^2{\bf r'}\frac{\delta n({\bf r'})}{\sqrt{|{\bf r}-{\bf r'}|^2+d^2}},
\end{equation}
where ${\bf r}$ is the position vector in the heterostructure plane, $\delta n({\bf r})=n({\bf r})-n_0$ is the donor density fluctuation in the doping plane with correlations $\overline{\delta n({\bf r})\delta n({\bf r}')}=n_0\delta({\bf r}-{\bf r}')$, $n_0$ being the mean concentration of the 2D electron system, and overlining represents averaging over disorder realizations. It is taken into account the screening of dopant charges by negatively charged electrons. Therefore we take $n_0$ as the effective density of donors. Computing the derivative of $\delta \varphi({\bf r},d)$ with respect to $d$ leads to the electric field fluctuation
$
\delta{\mathcal E}({\bf r},d)=\frac{ed}{4\pi\varepsilon_0\varepsilon}\int d^2{\bf r'}\frac{\delta n({\bf r'})}{(|{\bf r}-{\bf r'}|^2+d^2)^{3/2}}.
$
The correlator evaluates as $\overline{\delta{\mathcal E}({\bf r_1},d)\delta{\mathcal E}({\bf r_2},d)}=\left(\frac{e}{2\varepsilon_0\varepsilon}\right)^{2}n_0\tilde{\delta}({\bf r_1},{\bf r_2})$, with
\begin{equation}
\tilde{\delta}({\bf r_1},{\bf r_2})=\frac{d^2}{4\pi^2}\int dxdy\,\frac{1}{((x_1-x)^2+(y_1-y)^2+d^2)^{3/2}((x_2-x)^2+(y_2-y)^2+d^2)^{3/2}}.
\end{equation}
Here $x_{1/2}$, $y_{1/2}$ are the coordinates of ${\bf r_{1/2}}$. $\tilde{\delta}({\bf r_1},{\bf r_2})$ satisfies the properties $\int d^2{\bf r_1}\,\tilde{\delta}({\bf r_1},{\bf r_2})=1$, $\tilde{\delta}({\bf r_1},{\bf r_2})\to 0$ as $d\to0$ if $|{\bf r_1}-{\bf r_2}|\neq 0$ and $\tilde{\delta}({\bf r_1},{\bf r_2})\to\infty$ as $d\to0$ if $|{\bf r_1}-{\bf r_2}|=0$. So it realizes the Dirac delta function when $d$ is the smallest length scale.

Let us consider the quantum mechanical average
$
\langle \delta{\mathcal E}\rangle=\int d^2{\bf r}\,\delta{\mathcal E}({\bf r},d)\psi({\bf r})^2,
$
$\psi({\bf r})=\exp(-|{\bf r}|^2/4\ell^2)/\sqrt{2\pi}\ell$ being the ground state wavefunction with characteristic length $\ell=\sqrt{\hbar/2m_G\omega_0}$. From the above relations we obtain the field fluctuation:
\begin{equation}
\overline{\langle\delta{\mathcal E}\rangle^2}=\left(\frac{e}{2\varepsilon_0\varepsilon}\right)^2n_0\int d^2{\bf r}\,\psi({\bf r})^4=\left(\frac{en_0^{1/2}}{4\sqrt{\pi}\varepsilon_0\varepsilon\ell}\right)^2.
\end{equation}
As a consequence in the regime $d<\ell<D$ ($D$ being the separation between the tunnel gates) the typical dot-to-dot g-factor variation due to potential disorder is simply computed:
\begin{equation}
\delta g_{\rm eff}=\frac{\delta E_{z0}}{E_{z0}}\Big(\frac{3}{2}g_{\rm eff}^{(1)}+\frac{1}{2}g_{\rm eff}^{(2)}+g_{\rm eff}^{(3)}\Big)=\frac{\delta\mathcal{E}}{\mathcal{E}}\Big(g_{\rm eff}^{(1)}+\frac{1}{3}g_{\rm eff}^{(2)}+\frac{2}{3}g_{\rm eff}^{(3)}\Big)=\frac{1}{4(\pi n_0)^{1/2}\ell}\Big(g_{\rm eff}^{(1)}+\frac{1}{3}g_{\rm eff}^{(2)}+\frac{2}{3}g_{\rm eff}^{(3)}\Big).
\end{equation}

\section{Summary of electrostatic gate tunings for the quadruple dot device}

Values of the resonance frequencies for dot 1 and voltages applied to electrostatic gates in the quadruple dot device are given in Table. \ref{tab:example_resonance_extra} for different tuning configurations. This corresponds to the resonance lines shown in Fig. 3 of the main text.

\begin{table}[h!]
	\begin{tabular}{ | c | c | c | c | c | c | c |}
		\hline
		Quantity/Parameter & Mean & Tuning \textbf{A} & Tuning \textbf{B} & Tuning \textbf{C} & Tuning \textbf{D} & Tuning \textbf{E} \\ \hline
		Resonance frequency [GHz] & 15.624 & -0.144 & -0.084 & -0.024 & +0.096 & +0.156 \\ \hline
		\hline
		$V_{\rm D1}$ [mV] & -395.0	& +0.0 & -15.0 & +5.0 & +5.0 & +5.0 \\ \hline
		$V_{\rm D2}$ [mV] & -477.0 & -3.0 & +2.0 & -3.0 & +2.0 & +2.0 \\ \hline
		$V_{\rm D3}$ [mV] & -260.0 & -1.0 & -1.0 & -6.0 & +4.0 & +4.0 \\ \hline
		$V_{\rm P1}$ [mV] & 257.0 & -1.6 & +14.9 & -4.9 & -5.2 & -3.2 \\ \hline
		$V_{\rm P2}$ [mV] & 75.8 & +2.3 & +2.1 & -1.2 & -1.6 & -1.6 \\ \hline
		$V_{\rm P3}$ [mV] & -92.1 & +2.9 & +1.9 & +8.5 & -6.6 & -6.6 \\ \hline
		$V_{\rm P4}$ [mV] & 33.7 & +6.6 & -2.9 & +3.9 & -3.8 & -4.0 \\ \hline
		$V_{\rm SD1b}$ [mV] & -246.9 & -0.6 & +0.4 & -0.6 & +0.4 & +0.4 \\ \hline
		$V_{\rm SD2b}$ [mV] & -31.4 & -2.6 & +1.4 & -3.6 & +3.4 & +1.4 \\ \hline
	\end{tabular}
	\caption{Extracted resonance frequencies of dot~1 for five different gate voltage conditions (before subtraction of 22~MHz, which comes from the adiabatic inversion method). Conditions are ordered according to the frequency values, with maximum difference of 300~MHz between conditions A and E. We give the gate voltages that differ between the measurements. The accuracy of the resonance frequency is $\pm 20$~MHz, which includes $\pm 5$~MHz from the measurement resolution and $\pm 15$~MHz from the expected random nuclear spin distributions.}
	\label{tab:example_resonance_extra}
\end{table}

\section{EDSR data for different setups}

Below we show the plots of EDSR data collapse obtained similarly as in Fig. 2. The fitted parameters, uncertainties and dot-to-dot fluctuations are summarized in Table I of the main text. 

\begin{figure}[h!]
        \centering
        \begin{subfigure}[b]{0.475\textwidth}
            \centering
            \includegraphics[width=\textwidth]{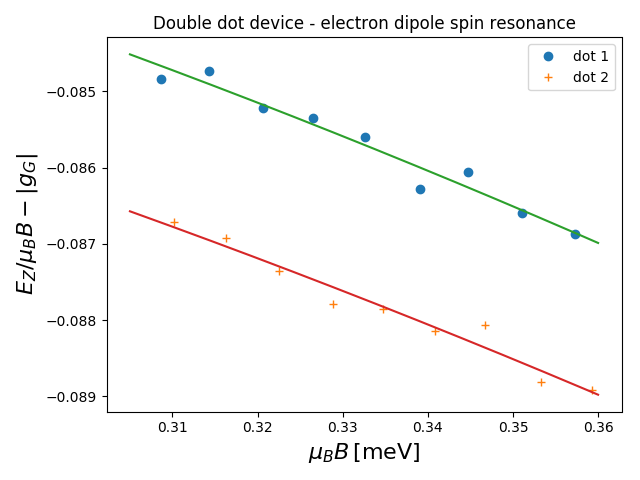}
        \end{subfigure}
        \hfill
        \begin{subfigure}[b]{0.475\textwidth}
            \centering
            \includegraphics[width=\textwidth]{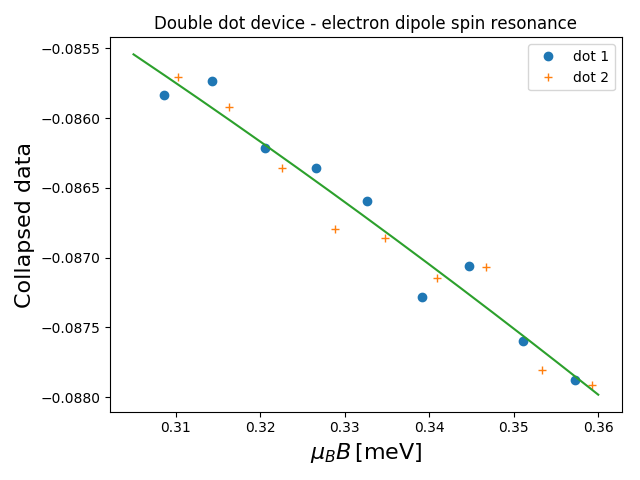}
            \label{collapsedouble}
        \end{subfigure}
        \hfill
        \begin{subfigure}[b]{0.475\textwidth}
            \centering
            \includegraphics[width=\textwidth]{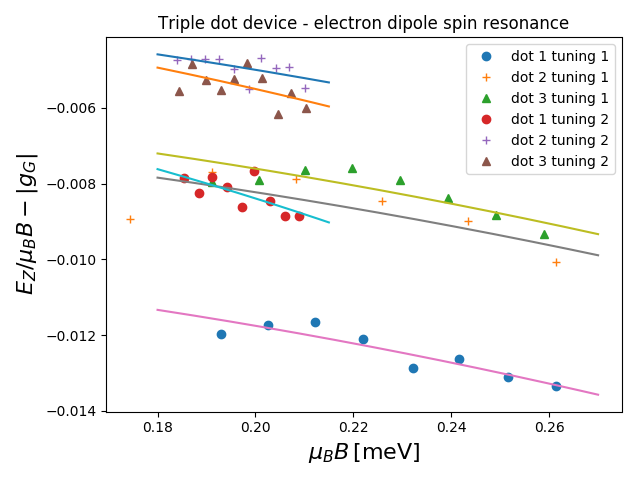}
        \end{subfigure}
        \hfill
        \begin{subfigure}[b]{0.475\textwidth}
            \centering 
            \includegraphics[width=\textwidth]{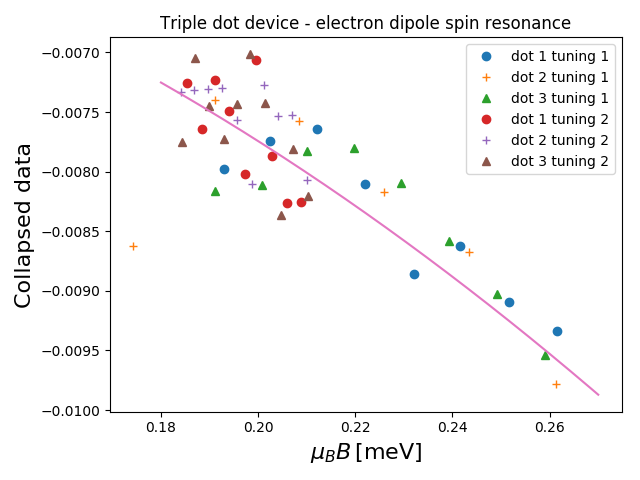}
            \label{collapsetriple}
        \end{subfigure}
        \caption{Experimental data for the double and triple dot devices. The bare data are shown on the left hand side and the collapsed data are on the right hand side. Data collapse is obtained as in Fig. 2 of the main text, where g-factors for the different dots are shifted by a constant value.}
        \label{collapse}
\end{figure}

\section{Temporal variation in resonance frequency}

\begin{figure}[!h]
	\includegraphics[width=0.8\textwidth]{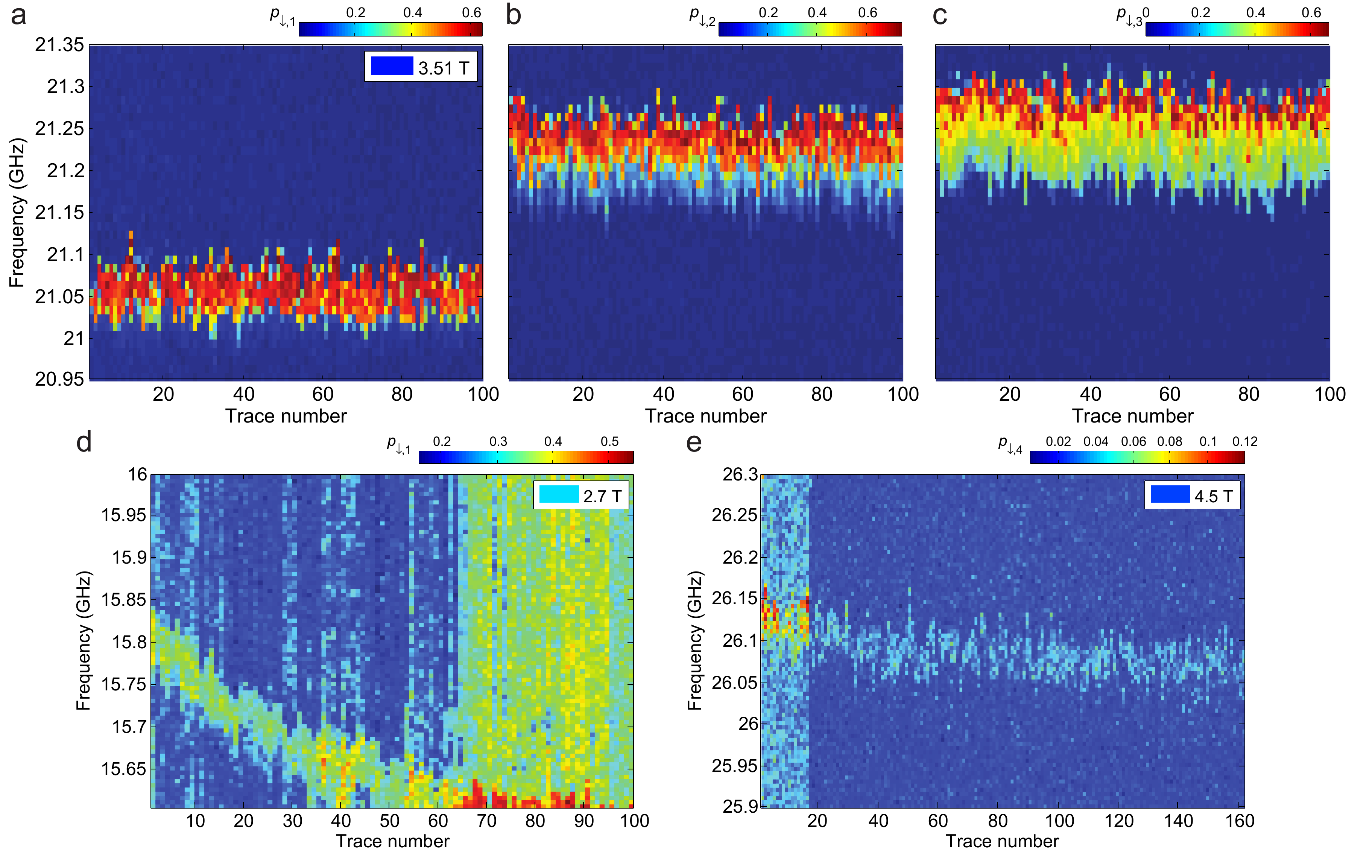}
	\caption{(a-e) Continuously measured resonance frequencies showing time dependence. Microwave resonance conditions give the Zeeman splitting. The frequency is swept from high to low values for each case. (a-c) Measured resonance on a triple quantum dot as in \cite{S_Baart2016a}. The total measurement time is 12 hours. (d,e) Measurement on a quadruple quantum dot as in \cite{S_Fujita2017}. (d) Down spin fraction of dot~1 is measured at $B$=2.7~T. Total measurement time is 8.5 hours, where a total shift of 200~MHz is observed. (e) Measurement on dot~4 at 4.5~T. Total measurement time is 22.8 hours.}
	\label{fig:EDSR_time_dependence}
\end{figure}

Figure~\ref{fig:EDSR_time_dependence} shows measurements where the resonance frequency of the spin in each dot changes with time while the microwave excitation is continuously measured. The total measurement times were on the order of hours and each frequency sweep was on the order of minutes. Although Figs.~ \ref {fig:EDSR_time_dependence}d,e indicated a monotonic frequency shift, the EDSR measurements in the main text were taken with only a few sweeps. Therefore we expect only a MHz order shift on the resonance frequency, and consequently these temporal shifts will not explain the gate voltage nor the dot dependent frequency shift shown in the main text. Additionally, there are smaller frequency fluctuations (about 5~MHz) on top of each sweep.

These slow and fast frequency variations presumably originate from the interaction with nuclear spins. The slower shift in Figs.~\ref{fig:EDSR_time_dependence}d,e could be due to dynamic-nuclear polarization~\cite{S_Chekhovich2013} similarly as in \cite{S_Vink2009}. The faster variation matches the observations of an electron spin interacting with a random distribution of nuclear spins~\cite{S_Bluhm2010b,S_Delbecq2016}.

\begin{thebibliography}{99}
\bibitem{S_commentRashba} Rigorously the electric field that appears in the term (6) is the effective electric field acting in the valence band \cite{S_Winkler03}. Taking this into account somewhat changes the contribution of this term to the Zeeman splitting but does not significantly change the main results.
\end{thebibliography}
\end{document}